\documentclass{article}

\usepackage{microtype}
\usepackage{graphicx}
\usepackage{subfigure}
\usepackage{booktabs} 

\usepackage{hyperref}

\usepackage[accepted]{icml2024}

\usepackage{amsmath}
\usepackage{amssymb}
\usepackage{mathtools}
\usepackage{amsthm}
\usepackage{multirow}
\usepackage{bm}
\usepackage{bbm}

\usepackage[capitalize,noabbrev]{cleveref}

\theoremstyle{plain}
\newtheorem{theorem}{Theorem}[section]

\newtheorem{lemma}[theorem]{Lemma}

\theoremstyle{definition}

\theoremstyle{remark}

\newcommand{\wwj}[1]{\textcolor{black}{#1}}
\newcommand{\wh}[1]{\textcolor{black}{#1}}
\newcommand{\myyn}[1]{\textcolor{black}{#1}}
\usepackage[textsize=tiny]{todonotes}

\newcommand{\eg}{\textit{e}.\textit{g}., }

\icmltitlerunning{
	\wh{Correcting Diffusion-Based Perceptual Image Compression with Privileged End-to-End Decoder}}

\begin{document}
	
	\twocolumn[
	\icmltitle{
		\wh{Correcting Diffusion-Based Perceptual Image Compression \\ with Privileged End-to-End Decoder}
	}

	
	
	\icmlsetsymbol{equal}{*}
	
	\begin{icmlauthorlist}
		\icmlauthor{Yiyang Ma}{wict}
		\icmlauthor{Wenhan Yang}{pcl}
		\icmlauthor{Jiaying Liu}{wict}
	\end{icmlauthorlist}
	
	\icmlaffiliation{wict}{Wangxuan Institute of Computer Technology, Peking University, Beijing, China}
	\icmlaffiliation{pcl}{Pengcheng Laboratory, Shenzhen, China}
	
	\icmlcorrespondingauthor{Yiyang Ma}{myy12769@pku.edu.cn}
	\icmlcorrespondingauthor{Jiaying Liu}{liujiaying@pku.edu.cn}
	
	\icmlkeywords{Machine Learning, ICML}
	
	\vskip 0.3in
	]
	
	
	
\printAffiliationsAndNotice{}  
 
\begin{abstract}
\wh{The images produced by diffusion models can attain excellent perceptual quality.
However, it is challenging for diffusion models to guarantee distortion,
hence the integration of diffusion models and image compression models still needs more comprehensive explorations.}
This paper presents a diffusion-based image compression method that employs a \wh{privileged} end-to-end decoder model as correction, \wh{which achieves better perceptual quality while guaranteeing the distortion to an extent.}
We build a diffusion model and design \wh{a novel paradigm that} combines the diffusion model and an end-to-end decoder, \wh{and the latter is responsible for transmitting the privileged information extracted at the encoder side.}
Specifically, we theoretically analyze the reconstruction process of the diffusion models at the encoder side with the original images being visible. 
Based on the analysis, we introduce an end-to-end \wh{convolutional} decoder to \wh{provide} a better approximation of the score function $\nabla_{\mathbf{x}_t}\log p(\mathbf{x}_t)$ at the \wh{encoder} side and \wh{effectively transmit the combination}.
Experiments \wh{demonstrate} the superiority of our method \wh{in} both distortion and perception compared with previous perceptual compression methods.
\end{abstract}

\begin{figure*}
    \centering
    \includegraphics[width=0.82\linewidth]{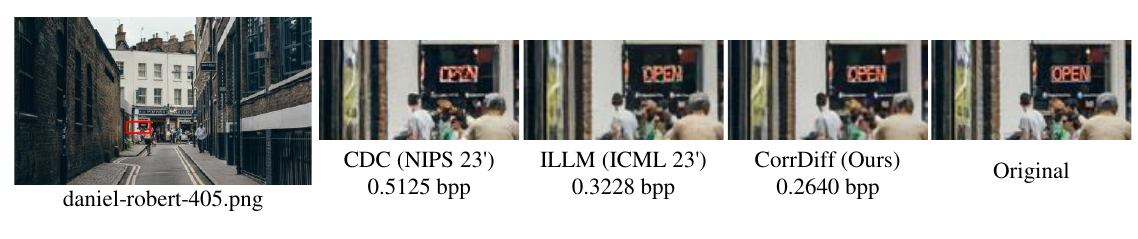}
    \vspace{-4mm}
    
    \caption{Visual results compared to CDC \cite{2023CDC} and ILLM \cite{2023ILLM}. The patch is cropped from \textit{daniel-robert-405.png} from CLIC \textit{professional} dataset \cite{2020CLIC}. \textbf{[Zoom in for best view]}}
    \vspace{-4mm}
    
    \label{fig: teaser}
\end{figure*}

\section{Introduction}
\label{sec: intro}
\wh{Image compression aims to minimize the amount of data required to represent the image while retaining as much relevant information as possible, to reconstruct the image with high fidelity.}
%
\wh{The persistent evolution of image compression technologies promotes the propensity of a series of emerging applications, 
\eg video streaming and augmented reality.}
\wh{With the rapid increase in existing image resolutions, \eg the emergence and polarity of High Definition, Full High Definition, and Ultra High Definition, 
image compression has drawn greater attention and interest.}

\wh{
In the past few decades, the conventional image compression pipeline consists of several fundamental modules: transformation, quantization, and entropy coding.
The images are first decomposed into less correlated components, whose distributions are then estimated and modeled.}
\wh{This framework leads to the birth and success of a series of coding standards, \eg JPEG \cite{1991JPEG} and BPG \cite{2018BPG}.}
\myyn{The compression method can also adaptively adjust by the corresponding input, achieving better-customized results \cite{2011SAO_HEVC, 2012SAO_HEVC}.}
\wh{However, persistent manual optimization and pattern expansion result in an overly complex framework, gradually revealing performance bottlenecks as development progresses.}

\wh{With the advancement of deep learning (DL), recent years have shown that DL-based image compression methods significantly surpass classical methods in balancing bit rate and reconstruction quality.}
\wh{In the beginning}, deep neural networks are employed \wh{to capture the nonlinear mapping relationship, to augment the function of existing modules of image codecs}~\cite{2017ICAE, 2017ICRNN, 2017RealTimeIC}.
%
\wh{
The later efforts~\cite{2017E2E, 2018Hyper, 2018HyperContext, 2015variableRI} make the entropy estimation learnable and lead compression techniques to enter the era of end-to-end training.
}

To evaluate the performance of image compression methods, there are two categories of metrics including distortion and perception \cite{2019DPTrade-off}. Distortion metrics (\eg PSNR, MS-SSIM~\cite{2003MS-SSIM}) which measure the fidelity of reconstructed images are leveraged by most image compression methods. Perception metrics refer to the subjective evaluation of human eyes.
\wh{However, when image compression methods reach a certain level of fidelity, rich evidence is provided to theoretically and experimentally prove that 
optimizing distortion inevitably leads to the degradation of perceptual quality 
~\cite{2019DPTrade-off, 2023ILLM}.}
Such degradation usually includes over-smoothness or blurring, which has minor effects on the distortion metrics \wh{but incurs a significant decrease in visual perception.}
%
\wh{To address the issue, the generation models, which are good at generating human visually pleasing details, are incorporated into the image compression methods for achieving better subjective visual quality.}
%
\citet{2019GAN_LIC_LowBit, 2020HiFiC} propose to employ Generative Adversarial Networks (GANs) \cite{2014GAN} as image decoders \wh{to generate the reconstruction results with rich details.}
\citet{2022PO-ELIC} and \citet{2023ILLM} improve the GAN-based methods with \wh{advanced perception models}.
Due to the great \wh{recent} success of diffusion-based models \cite{2019ScoreMatching, 2020DDPM}, \citet{2023CDC} leverage diffusion models to reconstruct images,
getting impressive results in terms of perception.
However, it has been validated that vanilla diffusion models tend to reconstruct images with richer visual details but less fidelity to the original images~\cite{2021SR3, 2023CDC}.
\wh{
Intuitively, the powerful detail reconstruction capacity arises from the progressive process of adding/removing noise, which might not be friendly to distortion measures.
Furthermore, the diffusion models rely on the efficacy of sampling, which further results in the difficulty of obtaining effective deterministic compression mapping.
}

%
%
To leverage \wh{both} the generation \wh{capacity of} diffusion models \wh{with less distortion loss},
\wh{
we propose to transmit a \textit{correction item} compactly to correct the sampling process of the diffusion decoder.
This correction is generated from the bitstream via an end-to-end convolutional decoder adaptively, 
which maintains the low bit rate while guaranteeing the distortion and improving the perceptual quality.}
%
%
\wh{In detail, we} first theoretically analyze the approximation error of the score function which is leveraged in the reconstruction process of the diffusion model by a score network. 
Then, we introduce a privileged end-to-end convolutional decoder and linearly combine \wh{such} decoder \wh{with} the score network \wh{via} a \wh{mathematically derived} factor to build an approximation of the \wh{above-mentioned error}.
At last, we can simply send these linear {factors that are used to combine the \wh{two components} with a few bits as \wh{privileged information},
\wh{assisting} the \wh{decoder} to correct the sampling process, which makes reconstruction results obtain improved visual quality.} The proposed method is called ``CorrDiff'' (abbreviated from ``Corrected Diffusion'').
Noting that the target of reconstructing images with high fidelity of the original images, comparatively in our work, we refer the concept of ``perceptual quality'' to the general superiority of a set of image-level perception-oriented metrics (\eg LPIPS \cite{2018LPIPS}) to evaluate image pair-wise fidelity in this paper.

The contributions can be summarized as follows:

\begin{enumerate}
    \item We propose a novel diffusion-based image compression framework, CorrDiff, with a \wh{privileged} end-to-end decoder.
    \wh{This privileged decoder helps correct the sampling process with only a few bits to facilitate the decoder side to achieve better reconstruction.}
    \item We theoretically analyze the sampling process of diffusion models and further derive the design of the end-to-end correction \wh{paradigm}.
    \item We conduct extensive experiments including diverse metrics and give ablation studies to \wh{demonstrate the superiority of the proposed image compression method as well as the effectiveness of each component.}
\end{enumerate}


\section{Related Works}

\subsection{Learned Image Compression Methods}

In recent years, \wwj{along with the development of deep learning}, more and more DL-based image compression methods have been proposed. \citet{2016GDN} 
design \wwj{generalized divisive normalization} which is widely used in the image compression task. \citet{2018Hyper} propose \wh{to use} a hyper-bit rate to represent the mean of the main bit rate, reducing the cost of encoding the bit rates.
\citet{2018HyperContext} further \wh{employ} a context model to predict the representation based on previous positions. \citet{2020DGML} introduce attention mechanism \cite{2017Attention} to handle the relation between different regions. 

To achieve better perceptual quality which is closer to human perception, \wh{a series of} methods leveraging generative models to build decoders are proposed \cite{2019GAN_LIC_LowBit, 2020HiFiC, 2022PO-ELIC, 2023MultiRealism, 2023ILLM, 2023CDC}.

\subsection{Generative Models}

As the name implies, \wwj{image} generative models aim at creating novel images \wh{that} contain visual details with high perceptual quality. \citet{2014GAN} propose GANs which contain a generator and a discriminator to compete. \citet{2014VAE} design variational auto encoder to explicitly model the posterior probability distribution of images. \citet{2018GLOW} propose flow-based models to map the distribution of images to a Gaussian distribution in an invertible way and generate images through the reverse map.

In the past few years, \citet{2020DDPM} introduce diffusion models in a simplified form. Diffusion models create images by gradually de-noising from \wh{the beginning} of a Gaussian noise, which \wh{has} been validated to have \wh{a great capacity} to create \wh{high-quality} \wwj{contents} \cite{2022Imagen, 2022StableDiffusion, 2023UMMDiffusion, 2022MMDiffusion}. The corresponding theories grow fast \cite{2019ScoreMatching, 2021ScoreModel, 2021VariationalDPM, 2021ImprovedDDPM, 2022AnalyticDPM, 2021ClassiferFreeGuidance}, making it mathematically complete. In low-level vision tasks, diffusion models also achieve impressive performance \cite{2023DiffIR, 2023BoostDiffSR}. Thus, it is intuitive to employ diffusion models in the task of image compression to reconstruct images with high perceptual quality. However, it is challenging because the integration of diffusion models and image compression models is non-trivial as we have mentioned in the abstract.

\subsection{Distortion and Perception Metrics}

To evaluate the performance of image compression methods, several metrics are employed. They can be divided into two categories, distortion and perception. In the category of distortion, PSNR is the most widely-used \wh{metric that measures} the mean square error (MSE). Furthermore, \citet{2003MS-SSIM} propose multi-scale structural similarity (MS-SSIM) to compare patch-level similarity between two images. \citet{2013GMSD} design \wwj{GMSD, which compares the gradient of two images.}

The perceptual metrics leverage features and their combinations of pre-trained neural networks. LPIPS~\cite{2018LPIPS} calculates the summation of MSE between a pyramid of deep features, which can employ different networks as \wh{the backbone}. \wwj{DISTS}~\cite{2020DISTS} transforms images through Gaussian kernels and \wh{evaluates} the transformed features. Unlike previous \wh{image-level} metrics, 
FID~\cite{2017TTUR&FID} \wwj{is a widely-used metric to measure the divergence between two distributions. It is commonly leveraged to evaluate the performance of image generation models, as these models aim to create novel images within the data distribution.}
However, it is not suitable enough to measure the performance of image compression models because the target of such models is to reconstruct the original images with high fidelity, instead of generating new images.

\begin{figure*}
    \centering
    \includegraphics[width = 0.85\linewidth]{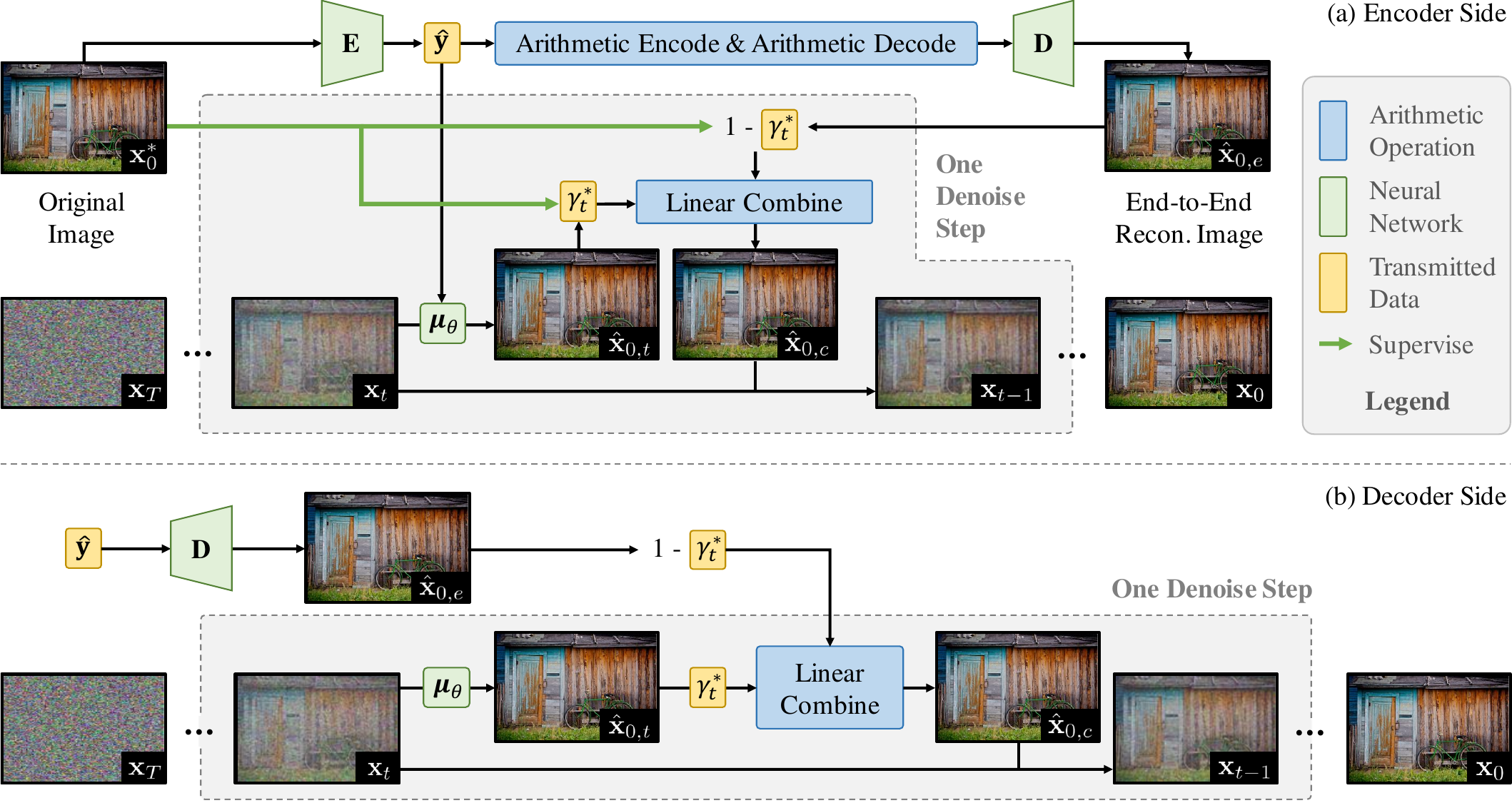}
    \vspace{-4mm}
    
    \caption{
    The framework of the proposed method. $\mathbf{E}$ denotes the encoder, $\mathbf{D}$ denotes the end-to-end decoder, and $\bm{\mu}_\theta$ denotes the score network. The yellow frame denotes the transmitted parts.
    The subimage (a) illustrates the pipeline at the encoder side, which obtains the representation $\hat{\mathbf{y}}$ and the factor set $\{\gamma^*_t\}_{t=1}^T$. The subimage (b) shows the \wh{reconstruction} process on \wh{the decoder} side.
    }
    \vspace{-4mm}
    
    \label{fig: method}
\end{figure*}

\section{Method}

We first review general theories of diffusion models, including the score-matching perspective and its corresponding differential equations. Then we analyze the approximation error of the score function by the score network when the original images are visible, which can provide privileged information and facilitate correcting the error at the decoder side.
Finally, we design a \wh{paradigm} to approximate the error through an external end-to-end decoder and send the approximation to the decoder side with a few bits. 
With the proposed correction \wh{mechanism}, we can achieve the goal of obtaining better reconstruction at the side of \wh{the decoder} \wh{in terms of} both distortion and perceptual quality. The proposed CorrDiff is illustrated in Fig.~\ref{fig: method}.

\subsection{Preliminaries}

Diffusion models \cite{2019ScoreMatching, 2020DDPM, 2021ScoreModel} are a kind of generative model \wh{that can} create impressively \wh{high-quality} images.
Diffusion models first perturb images $\mathbf{x}_0^*$ by adding Gaussian noise through a pre-specified distribution, then train a score function to estimate the noise injected into the images. 
By utilizing the score function iteratively, we can sample novel images from the distribution of pure Gaussian noise. The process of adding noise is called \textit{forward process} and the de-noising process is called \textit{reverse process}. The distribution of \textit{forward process} is given below:
\begin{align}
\label{eqn: forward process}
    q(\mathbf{x}_t|\mathbf{x}_0) &= \mathcal{N}(\mathbf{x}_t;\alpha(t)\mathbf{x}_0, \sigma^2(t)\mathbf{I}), \nonumber
    \\ 
    q(\mathbf{x}_T) &= \mathcal{N}(\mathbf{x}_T;\mathbf{0}, \mathbf{I}),
\end{align}
where $\alpha(t), \sigma(t)$ are differentiable hyper-parameter functions of $t$.
Furthermore, \citet{2021VariationalDPM, 2022DPM-Solver} prove that the following stochastic differential equation (SDE) has the same transition distribution with the conditional distribution $q(\mathbf{x}_t|\mathbf{x}_0)$ before at any $t \in [0, T]$:
\begin{equation}
    {\rm d}\mathbf{x}_t = f(t)\mathbf{x}_t{\rm d}t + g(t){\rm d}\mathbf{w}_t,
\end{equation}
where $\mathbf{w}_t$ is a standard Wiener process, and $f(t), g(t)$ are functions of $\alpha(t), \sigma(t)$, given by:
\begin{align}
    f(t) &= \frac{{\rm d}\log \alpha(t)}{{\rm d}t}, \nonumber
    \\
    g^2(t) &= \frac{{\rm d}\sigma^2(t)}{{\rm d}t} - 2\frac{{\rm d}\log \alpha(t)}{{\rm d}t}\sigma^2(t).
\end{align}
\citet{2021ScoreModel, 2022DPM-Solver} prove that the \textit{reverse process} can be done by solving the SDE below:
\begin{align}
\label{eqn: diffusion SDE}
    {\rm d}\mathbf{x}_t &= \big[f(t) \mathbf{x}_t - g^2(t)\nabla_{\mathbf{x}_t}\log q_t(\mathbf{x}_t)\big]{\rm d}t + g(t){\rm d}\bar{\mathbf{w}}_t, \nonumber
    \\
    \mathbf{x}_T &\sim \mathcal{N}(\mathbf{0}, \mathbf{I}),
\end{align}
where the score function $\nabla_{\mathbf{x}_t}\log q_t(\mathbf{x}_t)$ is estimated by a neural network. In practice, the network $\mathbf{s}_\theta(\mathbf{x}_t, t)$ is trained to estimate the scaled score function $-\sigma(t)\nabla_{\mathbf{x}_t}\log q_t(\mathbf{x}_t)$ by optimizing the loss function:
\begin{align}
    \mathcal{L} &= \mathbbm{E}_{t,\mathbf{x}_0}\big[\omega(t)\|\mathbf{s}_\theta(\mathbf{x}_t, t) + \sigma(t)\nabla_{\mathbf{x}_t}\log q_t(\mathbf{x}_t)\|^2\big] \nonumber
    \\
    &= \mathbbm{E}_{t,\mathbf{x}_0,\bm{\epsilon}}\big[\omega(t)\|\mathbf{s}_\theta(\mathbf{x}_t, t) - \bm{\epsilon} \|^2\big],
\end{align}
where $\omega(t)$ is the weighting function of loss terms of different $t$. Moreover, \citet{2021ScoreModel} \wh{gives} an ordinary differential equation which is the same \wh{as} the marginal distribution of Eqn.~\eqref{eqn: diffusion ODE}:
\begin{equation}
\label{eqn: diffusion ODE}
    \frac{{\rm d}\mathbf{x}_t}{{\rm d}t} = f(t) \mathbf{x}_t + \frac{g^2(t)}{2}\nabla_{\mathbf{x}_t}\log q_t(\mathbf{x}_t), \mathbf{x}_T \sim \mathcal{N}(\mathbf{0}, \mathbf{I}).
\end{equation}
After the training process, we can utilize the trained score network $\mathbf{s}_\theta$ and solve the Eqn.~\eqref{eqn: diffusion ODE} through numerical solvers like DPM-Solver \cite{2022DPM-Solver} or DDIM \cite{2020DDIM} to sample from the diffusion model. 

In this subsection, we discuss the general theories of diffusion models. In the following parts of this \wh{subsection}, we will present our method of leveraging diffusion models \wh{for} image compression \wh{via the proposed approach to correct} the \textit{reverse process} \wh{by taking the original images $\mathbf{x}_0^*$ as the available privileged information.} 

\subsection{Correcting the Score Function with Original Images}

As we have discussed in the previous subsection, we train a score network to estimate the score function of the diffusion model. When we manage to leverage diffusion models in the task of image compression, we first obtain the discretized image representation $\hat{\mathbf{y}}$ by an encoder $\mathbf{E}$ and further quantization following previous DL-based image compression methods \cite{2018Hyper, 2018HyperContext}:
\begin{equation}
    \hat{\mathbf{y}} = Q\big(\mathbf{E}(\mathbf{x}_0^*)\big),
\end{equation}
where we use the subscript $0$ to indicate noise-free images following the setting of diffusion models and the superscript $*$ to indicate original images.
We extend the score network $\mathbf{s}_\theta(\mathbf{x}_t, t)$ with $\hat{\mathbf{y}}$ as conditions $\mathbf{s}_\theta(\mathbf{x}_t, \hat{\mathbf{y}}, t)$. However, it is notable that the original images $\mathbf{x}_0^*$ are visible at the encoder side. It is intuitive to analyze how to correct the estimation of the conditioned score function $\nabla_{\mathbf{x}_t}\log q_t(\mathbf{x}_t|\hat{\mathbf{y}})$ at the encoder side with $\mathbf{x}_0^*$ as an additional condition. We have:
\begin{equation}
    q_t(\mathbf{x}_t|\hat{\mathbf{y}}, \mathbf{x}_0^*) = \frac{q_t(\mathbf{x}_0^*|\hat{\mathbf{y}}, \mathbf{x}_t) q_t(\mathbf{x}_t|\hat{\mathbf{y}})}{p(\mathbf{x}_0^*|\hat{\mathbf{y}})},
\end{equation}
through Bayes rule. Thus, we have:
\begin{align}
\label{eqn: modified score}
    \nabla_{\mathbf{x}_t}\log q_t(\mathbf{x}_t|\hat{\mathbf{y}}, \mathbf{x}_0^*) =& \nabla_{\mathbf{x}_t}\log q_t(\mathbf{x}_t|\hat{\mathbf{y}}) + \nonumber
    \\
    & \nabla_{\mathbf{x}_t}\log q_t(\mathbf{x}_0^*|\hat{\mathbf{y}}, \mathbf{x}_t),
\end{align}
because $p(\mathbf{x}_0^*|\hat{\mathbf{y}})$ is not related to $\mathbf{x}_t$. In Eqn.~\eqref{eqn: modified score}, it is noticed that the first item $\nabla_{\mathbf{x}_t}\log 
q_t(\mathbf{x}_t|\hat{\mathbf{y}})$ is estimated by the score function $\mathbf{s}_\theta(\mathbf{x}_t, \hat{\mathbf{y}}, t)$, and the second term $\nabla_{\mathbf{x}_t}\log q_t(\mathbf{x}_0^*|\hat{\mathbf{y}}, \mathbf{x}_t)$ is a correction item resulted from the visible original images $\mathbf{x}_0^*$. Therefore, if we can transmit such an item, we can assist the decoder side in reconstructing images with better quality. However, such correction items have the same dimensions as the original images. So, it is not effective to transmit them to the decoder side directly. Hence, we design a protocol to approximate the correction items through an external end-to-end decoder which only needs a few bits to send. We will state our protocol and its corresponding theories in the following subsection.
\begin{algorithm}[tb]
   \caption{Encoder Side with DDIM.}
   \label{alg: sender}
\begin{algorithmic}
   \STATE {\bfseries Input:} original image $\mathbf{x}_0^*$
   \STATE {\bfseries Require:} encoder $\mathbf{E}$,  score network $\bm{\mu}_\theta$, end-to-end decoder $\mathbf{D}$, hyper-parameter functions $\alpha(t), \sigma(t)$
   \STATE Initialize $\mathbf{x}_T \sim \mathcal{N}(\mathbf{0}, \mathbf{I})$
   \STATE $\hat{\mathbf{y}} \leftarrow Q(\mathbf{E}(\mathbf{x}_0^*))$
   \FOR{$t=T$ {\bfseries to} $1$}
        \STATE $\hat{\mathbf{x}}_{0, t} \leftarrow \bm{\mu}_{\theta}(\mathbf{x}_t, \hat{\mathbf{y}}, t)$
        \STATE $\hat{\mathbf{x}}_{0, e} \leftarrow \mathbf{D}(\hat{\mathbf{y}})$
        \STATE $\gamma^*_t \leftarrow \mathop{\arg\min}_\gamma \big[{\rm M}(\mathbf{x}_0^*, \gamma \hat{\mathbf{x}}_{0, t} + (1 - \gamma) \hat{\mathbf{x}}_{0, e})\big]$
        \STATE $\hat{\mathbf{x}}_{0, c} \leftarrow \gamma^*_t \hat{\mathbf{x}}_{0, t} + (1 - \gamma^*_t) \hat{\mathbf{x}}_{0, e}$
        \STATE $\hat{\bm{\epsilon}}_t \leftarrow \frac{1}{\sigma(t)}(\mathbf{x}_t - \alpha(t)\hat{\mathbf{x}}_{0, c})$
        \STATE $\mathbf{x}_{t-1} \leftarrow \alpha(t-1)\hat{\mathbf{x}}_{0, c} + \sigma(t - 1)\hat{\bm{\epsilon}}_t$
   \ENDFOR
   \STATE {\bfseries Send:} representation $\hat{\mathbf{y}}$, linear factors $\{\gamma^*_t\}_{t=1}^T$
\end{algorithmic}
\vspace{-0.5mm}
\end{algorithm}
\subsection{Approximation of the Correction via an External End-to-End Decoder}

First, as the score network actually estimates the noise injected into the original images, we can directly build pseudo noise-free images $\hat{\mathbf{x}}_{0, t}$ at any time-step $t$ following \cite{2020DDPM, 2023DPS}:
\begin{align}
\label{eqn: pseudo x0}
    \hat{\mathbf{x}}_{0, t} &:= \mathbb{E}_{\mathbf{x}_0\sim q_t(\mathbf{x}_0|\mathbf{x}_t, \hat{\mathbf{y}})}[\mathbf{x}_0|\mathbf{x}_t, \hat{\mathbf{y}}] \nonumber
    \\
    &= \frac{1}{\alpha(t)}\big(\mathbf{x}_t + \sigma^2(t)\nabla_{\mathbf{x}_t}\log p_t(\mathbf{x}_t|\hat{\mathbf{y}})\big).
\end{align}
Then, we can prove that (refer to Appendix Sec.~\ref{sec: sup proof}):
\begin{theorem}
\label{theo: transfer condition}
    The conditional distribution $q_t(\mathbf{x}_0^*|\hat{\mathbf{y}}, \mathbf{x}_t)$ can be approximated by $q_t(\mathbf{x}_0^*|\hat{\mathbf{y}}, \hat{\mathbf{x}}_{0, t})$.
\end{theorem}
\myyn{When given the pseudo noise-free image $\hat{\mathbf{x}}_{0, t}$, the $\mathbf{x}^*$ will distribute around $\hat{\mathbf{x}}_{0, t}$ which is less relative to the representation $\hat{\mathbf{y}}$. Thus, we have:}
\begin{equation}
\label{eqn: drop y_hat}
    q_t(\mathbf{x}_0^*|\hat{\mathbf{y}}, \hat{\mathbf{x}}_{0, t}) \approx q_t(\mathbf{x}_0^*| \hat{\mathbf{x}}_{0, t}).
\end{equation}
In the following derivation, we agree that images $\mathbf{x}_0^*$ are vectors (which are actually matrices) for formal simplicity. The exact shapes do not affect the correctness of the theories. With the approximation of \cref{theo: transfer condition} and Eqn.~\eqref{eqn: drop y_hat}, we transform the correction item by:
\begin{equation}
\label{eqn: replace condition}
    \nabla_{\mathbf{x}_t}\log q_t(\mathbf{x}_0^*|\hat{\mathbf{y}}, \mathbf{x}_t) \approx \nabla_{\mathbf{x}_t}\log q_t(\mathbf{x}_0^*|\hat{\mathbf{x}}_{0, t}).
\end{equation}
We have the chain rule of vector differentiation:
\begin{equation}
    \nabla_{\mathbf{x}_t}\log q_t(\mathbf{x}_0^*|\hat{\mathbf{x}}_{0, t}) = \bigg(\frac{\partial\hat{\mathbf{x}}_{0, t}}{\partial\mathbf{x}_t}\bigg)^\top\nabla_{\hat{\mathbf{x}}_{0, t}}\log q_t(\mathbf{x}_0^*|\hat{\mathbf{x}}_{0, t}).
\end{equation}
It is noticed that:
\begin{align}
    &(\nabla_{\mathbf{x}_t}\log q_t(\mathbf{x}_0^*|\hat{\mathbf{x}}_{0, t}))^\top \nabla_{\hat{\mathbf{x}}_{0, t}}\log q_t(\mathbf{x}_0^*|\hat{\mathbf{x}}_{0, t}) \nonumber
    \\
    =& \bigg[\bigg(\frac{\partial\hat{\mathbf{x}}_{0, t}}{\partial\mathbf{x}_t}\bigg)^\top\nabla_{\hat{\mathbf{x}}_{0, t}}\log q_t(\mathbf{x}_0^*|\hat{\mathbf{x}}_{0, t})\bigg]^\top \nabla_{\hat{\mathbf{x}}_{0, t}}\log q_t(\mathbf{x}_0^*|\hat{\mathbf{x}}_{0, t}) 
    \nonumber
    \\
    =& \nabla_{\hat{\mathbf{x}}_{0, t}}\log q_t(\mathbf{x}_0^*|\hat{\mathbf{x}}_{0, t})^\top \frac{\partial\hat{\mathbf{x}}_{0, t}}{\partial\mathbf{x}_t}\nabla_{\hat{\mathbf{x}}_{0, t}}\log q_t(\mathbf{x}_0^*|\hat{\mathbf{x}}_{0, t}).
\end{align}

\begin{algorithm}[tb]
   \caption{\wh{Decoder} Side with DDIM.}
   \label{alg: receiver}
\begin{algorithmic}
   \STATE {\bfseries Receive:} representation $\hat{\mathbf{y}}$, linear factors $\{\gamma^*_t\}_{t=1}^T$
   \STATE {\bfseries Require:} score network $\bm{\mu}_\theta$, end-to-end decoder $\mathbf{D}$, hyper-parameter functions $\alpha(t), \sigma(t)$
   \STATE Initialize $\mathbf{x}_T \sim \mathcal{N}(\mathbf{0}, \mathbf{I})$
   \FOR{$t=T$ {\bfseries to} $1$}
        \STATE $\hat{\mathbf{x}}_{0, t} \leftarrow \bm{\mu}_{\theta}(\mathbf{x}_t, \hat{\mathbf{y}}, t)$
        \STATE $\hat{\mathbf{x}}_{0, e} \leftarrow \mathbf{D}(\hat{\mathbf{y}})$
        \STATE $\hat{\mathbf{x}}_{0, c} \leftarrow \gamma^*_t \hat{\mathbf{x}}_{0, t} + (1 - \gamma^*_t) \hat{\mathbf{x}}_{0, e}$
        \STATE $\hat{\bm{\epsilon}}_t \leftarrow \frac{1}{\sigma(t)}(\mathbf{x}_t - \alpha(t)\hat{\mathbf{x}}_{0, c})$
        \STATE $\mathbf{x}_{t-1} \leftarrow \alpha(t-1)\hat{\mathbf{x}}_{0, c} + \sigma(t - 1)\hat{\bm{\epsilon}}_t$
   \ENDFOR
   \STATE {\bfseries Return:} reconstructed image $\mathbf{x}_0$
\end{algorithmic}
\vspace{-0.5mm}
\end{algorithm}

It is noticed that the pseudo noise-free images $\hat{\mathbf{x}}_{0, t}$ are obtained by the score network from $\mathbf{x}_t$ to estimate the original images $\mathbf{x}_0^*$. Considering that the original images $\mathbf{x}_0^*$ have positive correlation with $\mathbf{x}_t$ on average, we have:
\begin{equation}
    \mathbb{E}_{\mathbf{x}_t} \bigg[\frac{\partial\hat{\mathbf{x}}_{0, t}}{\partial\mathbf{x}_t}\bigg] \approx \mathbb{E}_{\mathbf{x}_t} \bigg[\frac{{\partial\mathbf{x}}_0^*}{\partial\mathbf{x}_t}\bigg] \succ \bm{0}.
\end{equation}
Thus we approximately have:
\begin{equation}
    \big[\nabla_{\mathbf{x}_t}\log q_t(\mathbf{x}_0^*|\hat{\mathbf{x}}_{0, t})\big]^\top \nabla_{\hat{\mathbf{x}}_{0, t}}\log q_t(\mathbf{x}_0^*|\hat{\mathbf{x}}_{0, t}) > 0,
\end{equation}
which indicates $-\nabla_{\hat{\mathbf{x}}_{0, t}}\log q_t(\mathbf{x}_0^*|\hat{\mathbf{x}}_{0, t})$ is a descent direction of the function $\log q_t(\mathbf{x}_0^*|\hat{\mathbf{x}}_{0, t}))$ of $\mathbf{x}_t$. Hence, we can leverage $\nabla_{\hat{\mathbf{x}}_{0, t}}\log q_t(\mathbf{x}_0^*|\hat{\mathbf{x}}_{0, t})$ as an approximation of the direction of $\nabla_{\mathbf{x}_t}\log q_t(\mathbf{x}_0^*|\hat{\mathbf{x}}_{0, t})$. Considering the relation between $\mathbf{x}_t$ and $\hat{\mathbf{x}}_{0, t}$ defined in Eqn.~\ref{eqn: pseudo x0}, we multiply $\nabla_{\hat{\mathbf{x}}_{0, t}}\log q_t(\mathbf{x}_0^*|\hat{\mathbf{x}}_{0, t})$ with the reciprocal of the factor $\frac{\sigma^2(t)}{\alpha(t)}$ to ensure similar numerical scale:
\begin{equation}
    \label{eqn: x_0 grad approx}
    \nabla_{\mathbf{x}_t}\log q_t(\mathbf{x}_0^*|\hat{\mathbf{x}}_{0, t}) \approx \frac{\alpha(t)}{\sigma^2(t)} \nabla_{\hat{\mathbf{x}}_{0, t}}\log q_t(\mathbf{x}_0^*|\hat{\mathbf{x}}_{0, t}).
\end{equation}
It is noticed that $\log q_t(\mathbf{x}_0^*|\hat{\mathbf{x}}_{0, t})$ describes the similarity between the original image $\mathbf{x}_0^*$ and the pseudo noise-free image $\hat{\mathbf{x}}_{0, t}$. Thus, in consideration of the perceptual 
characteristic of images, we utilize a perception-oriented metric ${\rm M}(\cdot, \cdot)$ which measures the distance of two images to estimate $\log q_t(\mathbf{x}_0^*|\hat{\mathbf{x}}_{0, t})$:
\begin{equation}
\label{eqn: M approx}
    \nabla_{\hat{\mathbf{x}}_{0, t}} \log q_t(\mathbf{x}_0^*|\hat{\mathbf{x}}_{0, t}) \approx - \nabla_{\hat{\mathbf{x}}_{0, t}}{\rm M}(\mathbf{x}_0^*, \hat{\mathbf{x}}_{0, t}).
\end{equation}
However, transmitting such an item is still not efficient. 
To give a further approximation with a few bits, we introduce \wh{an} end-to-end decoder $\mathbf{D}$ which directly decodes the representation $\hat{\mathbf{y}}$ to images $\hat{\mathbf{x}}_{0, e} = \mathbf{D}(\hat{\mathbf{y}})$ (we use the subscript $0$ to indicate noise-free images and $e$ to indicate end-to-end results). 
We approximate $- \nabla_{\hat{\mathbf{x}}_{0, t}}{\rm M}(\mathbf{x}_0^*, \hat{\mathbf{x}}_{0, t})$ on the direction of $\hat{\mathbf{x}}_{0, e} - \hat{\mathbf{x}}_{0, t}$. The accuracy of such an approximation depends on the accuracy of $\hat{\mathbf{x}}_{0, e}$, which is the result of end-to-end decoder $\mathbf{D}$. If the model $\mathbf{D}$ is well trained, the quality of $\hat{\mathbf{x}}_{0, e}$ can be guaranteed. The approximation is:
\begin{equation}
\label{eqn: linear combination}
    - \nabla_{\hat{\mathbf{x}}_{0, t}}{\rm M}(\mathbf{x}_0^*, \hat{\mathbf{x}}_{0, t}) \approx \big[\gamma^*_t \hat{\mathbf{x}}_{0, t} + (1 - \gamma^*_t) \hat{\mathbf{x}}_{0, e} \big] - \hat{\mathbf{x}}_{0, t},
\end{equation}
where the linear factor $\gamma^*_t$ is defined as:
\begin{equation}
    \gamma^*_t := \mathop{\arg\min}_{\gamma} \big[{\rm M}(\mathbf{x}_0^*, \gamma \hat{\mathbf{x}}_{0, t} + (1 - \gamma) \hat{\mathbf{x}}_{0, e})\big],
\end{equation}
to ensure that the combination is closer to the original image than $\hat{\mathbf{x}}_{0, t}$. As long as $\gamma^*_t \neq 1$, the difference $[\gamma^*_t \hat{\mathbf{x}}_{0, t} + (1 - \gamma^*_t) \hat{\mathbf{x}}_{0, e}] - \hat{\mathbf{x}}_{0, t}$ will be a descent direction of ${\rm M}(\mathbf{x}_0^*, \hat{\mathbf{x}}_{0, t})$. The linear factor $\gamma^*_t$ is just a float number, which is effortless to transmit. Such a $\gamma^*_t$ is quite easy to obtain by gradient descent through ${\rm M}(\cdot, \cdot)$ at every time-step $t$.
%
%
The implementations of ${\rm M}(\cdot, \cdot)$ can be variable (\eg LPIPS \cite{2018LPIPS} and DISTS \cite{2020DISTS}). 
In summary, taking Eqn.~\eqref{eqn: pseudo x0},~\eqref{eqn: replace condition},~\eqref{eqn: x_0 grad approx},~\eqref{eqn: M approx},~\eqref{eqn: linear combination} into~\eqref{eqn: modified score}, the corrected score function is given below (the corresponding proof refers to Appendix Sec.~\ref{sec: sup proof}):
\begin{theorem}
$\nabla_{\mathbf{x}_t}\log q_t(\mathbf{x}_t|\hat{\mathbf{y}}, \mathbf{x}_0^*)$ can be approximated by the following combination:
    \begin{align}
    \label{eqn: corrected score}
    &\nabla_{\mathbf{x}_t}\log q_t(\mathbf{x}_t|\hat{\mathbf{y}}, \mathbf{x}_0^*) 
    \nonumber
    \\
    \approx& \frac{\alpha(t)}{\sigma^2(t)}\big[\gamma^*_t \hat{\mathbf{x}}_{0, t} + (1 - \gamma^*_t) \hat{\mathbf{x}}_{0, e} \big]
    -\frac{\mathbf{x}_t}{\sigma^2(t)}.
    \end{align}
\end{theorem}
Furthermore, considering the characteristic of \wh{the} image compression task, we train the score network $\bm{\mu}_\theta(\mathbf{x}_t, \hat{\mathbf{y}}, t)$ to predict $\hat{\mathbf{x}}_{0, t}$ directly instead of training $\mathbf{s}_\theta(\mathbf{x}_t, \hat{\mathbf{y}}, t)$ to predict $\bm{\epsilon}$. Thus, the actual used score function is:
\begin{align}
\label{eqn: final score}
    &\nabla_{\mathbf{x}_t}\log q_t(\mathbf{x}_t|\hat{\mathbf{y}}, \mathbf{x}_0^*) 
    \nonumber
    \\
    \approx& \frac{\alpha(t)}{\sigma^2(t)}\big[\gamma^*_t \bm{\mu}_\theta(\mathbf{x}_t, \hat{\mathbf{y}}, t) + (1 - \gamma^*_t) \mathbf{D}(\hat{\mathbf{y}}) \big]
    -\frac{\mathbf{x}_t}{\sigma^2(t)}.
\end{align}
In conclusion, we depict the protocol below:
\begin{enumerate}
    \item \textbf{At the \wh{encoder} side}, \wh{we} solve the reverse process Eqn.~\eqref{eqn: diffusion ODE} with corrected score function Eqn.~\eqref{eqn: final score} at the encoder side and calculate the linear factors $\gamma^*_t$ at each time-step $t$ with the original images $\mathbf{x}_0^*$ being visible. The corresponding algorithm is shown in \cref{alg: sender}.
    \item \textbf{\wh{Sending}} the representation $\hat{\mathbf{y}}$ and the set of linear factors $\{\gamma^*_t\}_{t=1}^T$ to the \wh{decoder}.
    \item \textbf{At the \wh{decoder} side}, \wh{we} leverage the received representation $\hat{\mathbf{y}}$ and the set $\{\gamma^*_t\}_{t=1}^T$, \wwj{and use} the corrected score function Eqn.~\eqref{eqn: final score} to reconstruct the images.
    The corresponding algorithm is shown in \cref{alg: receiver}.
\end{enumerate}

\subsection{Model Training}

As we have depicted in previous subsections, our framework contains three main parts: 
the encoder $\mathbf{E}$, the end-to-end decoder $\mathbf{D}$ and the score network $\bm{\mu}_\theta$. 
Besides, being similar to previous image compression methods, our framework also contains an entropy model to predict the mean and variance of the representation $\hat{\mathbf{y}}$ to arithmetically encode and decode $\hat{\mathbf{y}}$ and estimate the bit rate during training. We train the model in two phases. 
First, we load the parameters of $\mathbf{E}$ and the entropy model from a pre-trained end-to-end image compression model and only train the score network $\bm{\mu}_\theta$ with the loss below:
\begin{align}
    \mathcal{L}_{\rm phase_1} = \mathbbm{E}_{t,\mathbf{x}_0,\bm{\epsilon}}\big[&\|\bm{\mu}_\theta(\mathbf{x}_t, \hat{\mathbf{y}}, t) - \mathbf{x}_0^* \|^2 + \nonumber
    \\
    & \lambda_{\mu} {\rm M}_{\mu} (\bm{\mu}_\theta(\mathbf{x}_t, \hat{\mathbf{y}}, t), \mathbf{x}_0^*) \big],
\end{align}
where ${\rm M}_{\mu}$ is the perceptual loss leveraged in the training process following previous perceptual image compression methods \cite{2020HiFiC, 2023CDC, 2023ILLM} and $\lambda_{\mu}$ is the corresponding loss weight. After the training process of only the score network $\bm{\mu}_\theta$,  we train the entire framework including $\mathbf{E}$, $\mathbf{D}$, the entropy model and $\bm{\mu}_\theta$ with the loss below:
\begin{align}
    \mathcal{L}_{\rm phase_2} = \ &\mathcal{L}_{\rm phase_1} + \|\mathbf{D}(\hat{\mathbf{y}}) - \mathbf{x}_0^*\|^2 + \nonumber
    \\
    &\lambda_{e}{\rm M}_e(\mathbf{D}(\hat{\mathbf{y}}), \mathbf{x}_0^*) + \lambda_{r}R(\hat{\mathbf{y}}),
\end{align}
where ${\rm M}_{e}$ is the perceptual loss of end-to-end decoder, $R(\cdot)$ is the estimated bit rate, and $\lambda_{e}$, $\lambda_{r}$ are the corresponding loss weights. The implementations will be given in Sec.~\ref{subsec: implementation}.

\begin{figure*}
    \centering
    \includegraphics[width=0.77\linewidth]{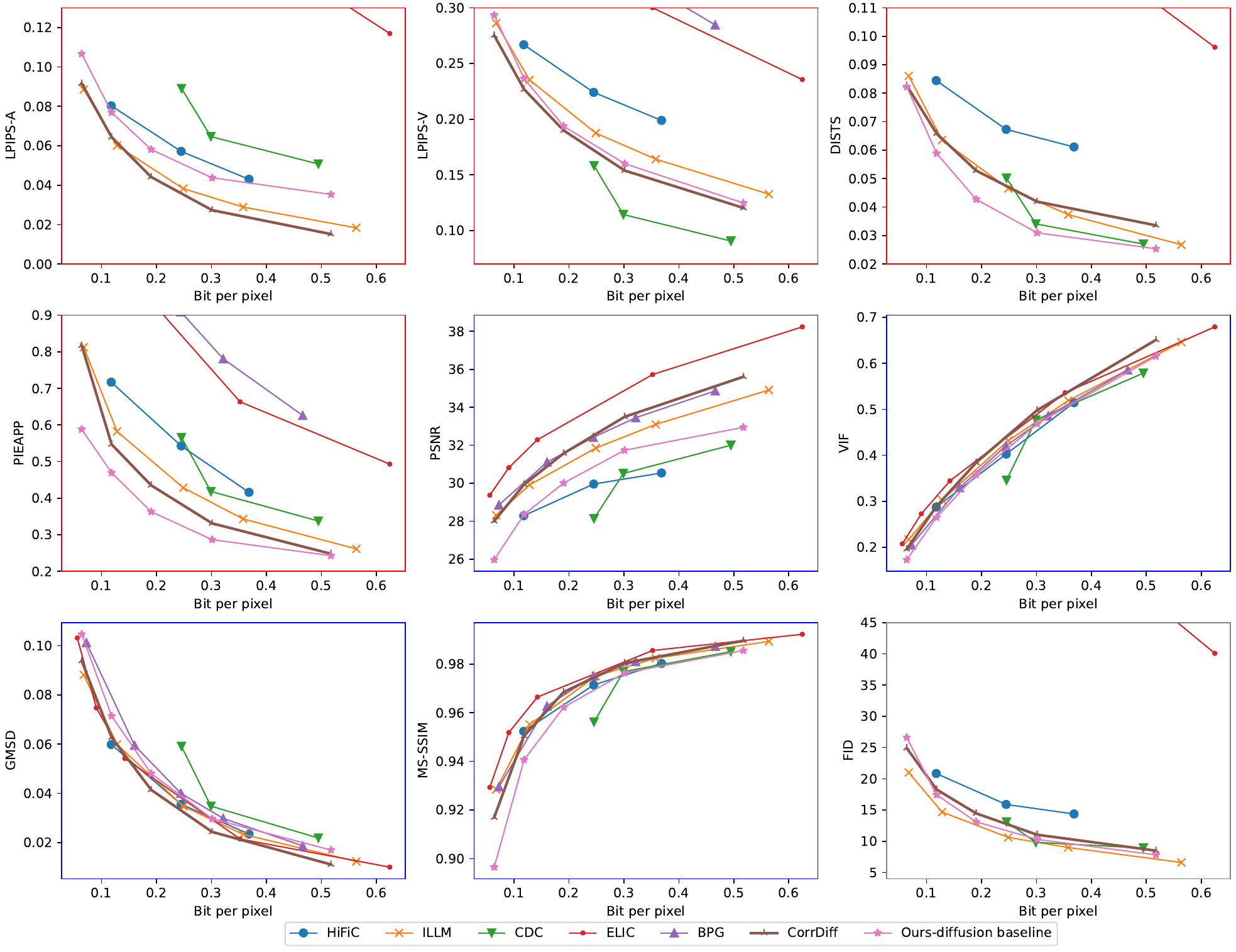}
    \vspace{-4mm}
    \caption{performance of diverse metrics on CLIC \textit{professional} dataset. \textbf{[Zoom in for best view]}}
    \vspace{-4mm}
    \label{fig: main rd}
\end{figure*}

\section{Experiments}

\subsection{Implementations}
\label{subsec: implementation}
\noindent \textbf{Model, Training and Inferring Settings.} We implement our score network based on the architecture of ADM \cite{2021ADM} with fewer parameters. \wwj{We leverage ELIC \cite{2022ELIC} including its encoder as $\mathbf{E}$, its entropy model and its decoder as our end-to-end decoder $\mathbf{D}$. Please refer to the supplementary materials for details.}

During training process, we utilize DISTS \cite{2020DISTS} as ${\rm M}_\mu$ and Alex-based \cite{2012Alex} LPIPS \cite{2018LPIPS} as ${\rm M}_e$. We use different implementations of ${\rm M}_\mu$ and ${\rm M}_e$ to avoid overfitting on one single metric. The perceptual weights are $\lambda_{\mu} = 0.16$ and $\lambda_e = 0.64$.
We use $5$ different bit rate weights $\lambda_r$ including $[0.5, 0.2, 0.1, 0.05, 0.02]$ to train $5$ models with different bit rates. We first train only the score network for $400, 000$ iterations and then train the entire framework for another $400, 000$ iterations with \wh{a} batch size of $8$, learning rate of $5e-5$ and optimizer of Adam \cite{2014Adam}. We train all the models on the dataset of DIV2K \cite{2017DIV2K} which includes $800$ \wh{high-resolution} images.
We randomly crop them into $256\times256$ patches in the training process. We also employ EMA with the rate of $0.9999$ to stabilize the training process following previous diffusion-based methods.

During inference, we use DDIM \cite{2020DDIM} as the diffusion sampler with $8$ steps and transmit $\gamma^*_t$ in the form of \texttt{float16} with $16$ bits, which means only $128$ bits in total. We leverage VGG-based \cite{2014VGG} LPIPS as $\rm M$ and use the vanilla gradient descent method provided by PyTorch \cite{2023PyTorch} to obtain all the $\gamma^*_t$ at every time-step.

\begin{figure*}
    \centering
    \includegraphics[width=0.78\linewidth]{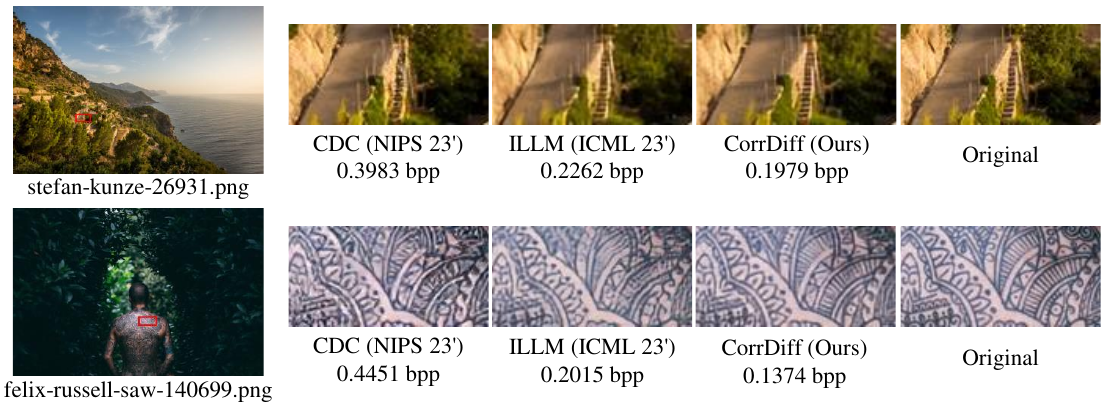}
    \vspace{-4mm}
    
    \caption{Visual results compared to CDC \cite{2023CDC} and ILLM \cite{2023ILLM}. \textbf{[Zoom in for best view]}}

    \vspace{-4mm}
    \label{fig: main visual}
\end{figure*}

\noindent \textbf{Datasets and Metrics.} 
We evaluate our method on $3$ datasets: Kodak \cite{1993Kodak}, CLIC \textit{professional} \cite{2020CLIC} and DIV2K-test \cite{2017DIV2K}. Kodak which includes $24$ images is one of the most widely-used datasets in the image compression task. CLIC \textit{professional} and DIV2K-test which contain $41$ and $100$ high-resolution images respectively are utilized to demonstrate the performance on large images.

To demonstrate the superiority of our method in both distortion and perceptual quality, we leverage a set of diverse metrics. For distortion, we utilize PSNR, VIF \cite{2006VIF} and GMSD \cite{2013GMSD}. For perception, we utilize LPIPS \cite{2018LPIPS}, DISTS \cite{2020DISTS}, and PIEAPP \cite{2018PIEAPP}. 
It is noticed that LPIPS can be employed with different backbones including AlexNet \cite{2012Alex} and VGG \cite{2014VGG}. Furthermore, we also leverage FID \cite{2017TTUR&FID}, which measures the distance between two image distributions, to evaluate the general quality of our reconstructed images. 
In Fig.~\ref{fig: main rd} which shows quantitative results, the charts with \textcolor{red}{red} frames are perceptual metrics, and curves with \textcolor{blue}{blue} frames are distortion metrics. FID is shown in a \textcolor{gray}{gray} frame due to its particularity.
%

\noindent \textbf{Baseline Methods.} We evaluate several image compression methods as baselines. We leverage BPG \cite{2018BPG} as a representative of classical methods. For DL-based methods, we employ an MSE-oriented method, ELIC (CVPR 2022) \cite{2022ELIC}, and a series of perceptual image compression methods including HiFiC (NeurIPS 2020) \cite{2020HiFiC}, ILLM (ICML 2023) \cite{2023ILLM} and CDC (NIPS 2023) \cite{2023CDC}.

\subsection{Quantitative Results}

We show the R-D curves of the $9$ metrics mentioned before on the dataset of CLIC \textit{professional} in Fig.~\ref{fig: main rd}. We further give results on the dataset of DIV2K and Kodak in Appendix Sec.~\ref{sec: rd on other dataset}. We also show the results of our method without the correction as an ablation study, which refer to Sec.~\ref{subsec: ablation}. \wh{It can be} seen that the proposed method achieves general superiority of diverse perceptual metrics along with better distortion \wh{compared with} other perceptual image compression methods, demonstrating the efficiency of the proposed method.
The only diffusion-based method, CDC, outperforms on \wh{LPIPS-VGG} because LPIPS-VGG is part of \wh{its} training target,
while \wh{it fails to achieve satisfactory results in} other metrics especially the distortion metrics (\eg PSNR, GMSD), revealing the limitation of vanilla diffusion-based image compression models.
ILLM, as the state-of-the-art perceptual image compression model, \wh{achieves competitive FID which is explicitly modeled during its training process}, but performs inferiorly \wh{in} all other metrics. Furthermore, FID is not a suitable enough metric to evaluate the performance of image compression methods because it measures the distance between two image distributions but the task of image compression focuses more on the fidelity of reconstructing the original images themselves instead of the distribution of all the reconstructed images. It is notable that, at the same time performing well on diverse perceptual metrics, our method also performs better on distortion metrics than previous perceptual image compression methods (\eg CDC, ILLM, and HiFiC), indicating that our method achieves a better distortion-perception trade-off.

\subsection{Quantitative Results}

To further demonstrate the perceptual quality of our results, we give several cases of different perceptual image compression methods in Fig.~\ref{fig: teaser} and Fig.~\ref{fig: main visual}. It is obvious that the 
reconstructed results of our method have more visual details with higher fidelity costing fewer bits. The full versions of the visual results refer to Appendix Sec. \ref{sec: sup visual}.

\subsection{Ablation Studies}
\label{subsec: ablation}

We \wh{conduct the ablation study on} the proposed design of introducing an external end-to-end decoder. 
We \wh{provide} performance of using only the diffusion part in Fig.~\ref{fig: main rd}.
We further give average BD rates \cite{2001BD-Rate} compared with the final setting on perception and distortion metrics of leveraging only the end-to-end part and using the score network $\bm{\mu}_\theta$ for one step to directly reconstruct the images in Tab.~\ref{tab: ablation}. When calculating the BD rates, we include FID as a perceptual metric and exclude the metrics that were leveraged as the targets during training (LPIPS-A for ``only end-to-end'' and DISTS for ``only diffusion'' and ``direct $\bm{\mu}_\theta$''). We further calculate the average BD rate of all the metrics.
As illustrated, leveraging only the diffusion can achieve fair perceptual results because \wh{it} is a powerful generative model, but leads to poor distortion being similar to CDC (which is a vanilla diffusion-based method).
Utilizing only the end-to-end decoder performs well in distortion as expected, but does not have the general superiority in diverse perceptual metrics, indicating the loss of perceptual quality. Employing only the score function has poor performance, revealing the significance of the diffusion model itself. The ablation studies prove that the proposed method \wh{can} achieve a better trade-off between distortion and perceptual quality.

\begin{table}[tb]
    \centering
    \caption{
    Average BD rates on perception and distortion metrics of different settings of our CorrDiff on CLIC \textit{professional} dataset.}
    \begin{tabular}{c|ccc}
    \toprule
         \multirow{2}{*}{Setting} & \multicolumn{3}{c}{BD Rate (\%)}
         \\
          & Perception & Distortion & Total
         \\
         \midrule
         CorrDiff (final) & - & - & -
         \\
         Only End-to-End & $+22.177$ & $-3.923$ & $+9.127$
         \\
         Only Diffusion & $+4.833$ & $+28.756$ & $+16.794$
         \\
         Direct $\bm{\mu}_\theta$ & $+12.113$ & $+10.358$ & $+11.235$
         \\
         \bottomrule
    \end{tabular}
    \vspace{-6.5mm}
    
    \label{tab: ablation}
\end{table}

\section{Conclusion}

In this paper, we propose a diffusion-based perceptual image compression method that leverages an \wh{priviliaged} end-to-end decoder to correct the score function. 
We leverage the fact that the original images are visible at the encoder side, propose the correction item and theoretically analyze the approximation of the correction which can be transmitted effectively. Experiments demonstrate the proposed method's superiority in terms of both distortion and perception. Further ablation studies validate the efficiency of the designed method of introducing correction of diffusion models.



\bibliography{AugDiffCo}
\bibliographystyle{icml2024}

\newpage
\appendix
\onecolumn
\section*{Appendix}
\setcounter{equation}{0}
\renewcommand\theequation{A.\arabic{equation}} 
\section{Proofs}
\label{sec: sup proof}
\begin{lemma} \cite{2023DPS}
    Let $\phi(\cdot)$ be \wh{a multivariate} Gaussian distribution with covariance matrix $\sigma^2\mathbf{I}$ being diagonal and mean $\bm{\mu}$. There exists a constant $L$, \textit{s.t.}, $\forall \mathbf{x}, \mathbf{y} \in \mathbbm{R}^d$,
    \begin{equation}
        |\phi(\mathbf{x}) - \phi(\mathbf{y})| \leq L \|\mathbf{x} - \mathbf{y}\|,
    \end{equation}
    where:
    \begin{equation}
        L = \frac{d}{\sqrt{2\pi}\sigma}e^{-1/2\sigma^2}.
    \end{equation}
\end{lemma}

\begin{theorem}
    The conditional distribution $q_t(\mathbf{x}_0^*|\hat{\mathbf{y}}, \mathbf{x}_t)$ can be approximated by $q_t(\mathbf{x}_0^*|\hat{\mathbf{y}}, \hat{\mathbf{x}}_{0, t})$.
\end{theorem}
\textit{Proof.} We have:
\begin{align}
    q_t(\mathbf{x}_0^*|\hat{\mathbf{y}}, \mathbf{x}_t) &= \int q_t(\mathbf{x}_0^*|\hat{\mathbf{y}}, \mathbf{x}_0) q_t(\mathbf{x}_0|\hat{\mathbf{y}}, \mathbf{x}_t){\rm d}\mathbf{x}_0 \nonumber
    \\
    &=\mathbbm{E}_{\mathbf{x}_0 \sim q_t(\mathbf{x}_0|\hat{\mathbf{y}}, \mathbf{x}_t)} [q_t(\mathbf{x}_0^*|\hat{\mathbf{y}}, \mathbf{x}_0)].
\end{align}
The distribution of $q_t(\mathbf{x}_0^*|\hat{\mathbf{y}}, \mathbf{x}_0)$ can be \wh{approximated} by a Gaussian distribution due to the characteristic of images:
\begin{equation}
    q_t(\mathbf{x}_0^*|\hat{\mathbf{y}}, \mathbf{x}_0) \approx \phi(\mathbf{x}_0).
\end{equation}
Thus we have the difference \wh{between} the two items:
\begin{align}
    &|q_t(\mathbf{x}_0^*|\hat{\mathbf{y}}, \mathbf{x}_t) - q_t(\mathbf{x}_0^*|\hat{\mathbf{y}}, \hat{\mathbf{x}}_{0, t})| \nonumber
    \\
    =&
    |\mathbbm{E}_{\mathbf{x}_0 \sim q_t(\mathbf{x}_0|\hat{\mathbf{y}}, \mathbf{x}_t)} [\phi(\mathbf{x}_0)] - \phi(\hat{\mathbf{x}}_{0, t})| \nonumber
    \\
    \leq&\int|\phi(\mathbf{x}_0) - \phi(\hat{\mathbf{x}}_{0, t})|{\rm d}Q(\mathbf{x}_0|\hat{\mathbf{y}}, \mathbf{x}_t) \nonumber
    \\
    \leq&\frac{d}{\sqrt{2\pi}\sigma}e^{-1/2\sigma^2}\int\|\mathbf{x}_0 - \hat{\mathbf{x}}_{0, t}\|{\rm d}Q(\mathbf{x}_0|\hat{\mathbf{y}}, \mathbf{x}_t),
\end{align}
and the item $\int\|\mathbf{x}_0 - \hat{\mathbf{x}}_{0, t}\|{\rm d}Q(\mathbf{x}_0|\hat{\mathbf{y}}, \mathbf{x}_t)$ is limited if the model is well-trained.
\hfill $\square$

The proof is inspired by \citet{2023DPS}.

\begin{theorem}
$\nabla_{\mathbf{x}_t}\log q_t(\mathbf{x}_t|\hat{\mathbf{y}}, \mathbf{x}_0^*)$ can be approximated by the following combination:
    \begin{align*}
    &\nabla_{\mathbf{x}_t}\log q_t(\mathbf{x}_t|\hat{\mathbf{y}}, \mathbf{x}_0^*) 
    \nonumber
    \\
    \approx& \frac{\alpha(t)}{\sigma^2(t)}\big[\gamma^*_t \hat{\mathbf{x}}_{0, t} + (1 - \gamma^*_t) \hat{\mathbf{x}}_{0, e} \big]
    -\frac{\mathbf{x}_t}{\sigma^2(t)}.
    \tag{\ref{eqn: corrected score}}
    \end{align*}
\end{theorem}
\textit{Proof.} Taking Eqn.~\eqref{eqn: pseudo x0},~\eqref{eqn: replace condition},~\eqref{eqn: x_0 grad approx},~\eqref{eqn: M approx},~\eqref{eqn: linear combination} into~\eqref{eqn: modified score}, we have:
\begin{align}
\label{eqn: implement}
    &\nabla_{\mathbf{x}_t}\log q_t(\mathbf{x}_t|\hat{\mathbf{y}}, \mathbf{x}_0^*) 
    \nonumber
    \\
    =&
    \nabla_{\mathbf{x}_t}\log q_t(\mathbf{x}_t|\hat{\mathbf{y}})  + \nabla_{\mathbf{x}_t}\log q_t(\mathbf{x}_0^*|\hat{\mathbf{y}}, \mathbf{x}_t)
    \nonumber
    \\
    \approx&
    \nabla_{\mathbf{x}_t}\log q_t(\mathbf{x}_t|\hat{\mathbf{y}}) + \nabla_{\mathbf{x}_t}\log q_t(\mathbf{x}_0^*|\hat{\mathbf{y}}, \hat{\mathbf{x}}_{0, t})
    \nonumber
    \\
    =&
    \nabla_{\mathbf{x}_t}\log q_t(\mathbf{x}_t|\hat{\mathbf{y}}) + \nabla_{\mathbf{x}_t}\log q_t(\mathbf{x}_0^*|\hat{\mathbf{x}}_{0, t})
    \nonumber
    \\
    \approx&
    \nabla_{\mathbf{x}_t}\log q_t(\mathbf{x}_t|\hat{\mathbf{y}}) + \frac{\alpha(t)}{\sigma^2(t)} \nabla_{\hat{\mathbf{x}}_{0, t}}\log q_t(\mathbf{x}_0^*|\hat{\mathbf{x}}_{0, t})
    \nonumber
    \\
    \approx&
    \nabla_{\mathbf{x}_t}\log q_t(\mathbf{x}_t|\hat{\mathbf{y}}) - \frac{\alpha(t)}{\sigma^2(t)} \nabla_{\hat{\mathbf{x}}_{0, t}}{\rm M}(\mathbf{x}_0^*, \hat{\mathbf{x}}_{0, t})
    \nonumber
    \\
    \approx&
    \nabla_{\mathbf{x}_t}\log q_t(\mathbf{x}_t|\hat{\mathbf{y}}) + \frac{\alpha(t)}{\sigma^2(t)}\big[(\gamma^*_t - 1)\hat{\mathbf{x}}_{0, t} + (1 - \gamma^*_t) \hat{\mathbf{x}}_{0, e} \big]
    \nonumber
    \\
    =&
    \frac{\alpha(t)}{\sigma^2(t)} \hat{\mathbf{x}}_{0, t} - \frac{\mathbf{x}_t}{\sigma^2(t)} + \frac{\alpha(t)}{\sigma^2(t)}\big[(\gamma^*_t - 1)\hat{\mathbf{x}}_{0, t} + (1 - \gamma^*_t) \hat{\mathbf{x}}_{0, e} \big]
    \nonumber
    \\
    =&
    \frac{\alpha(t)}{\sigma^2(t)}\big[\gamma^*_t \hat{\mathbf{x}}_{0, t} + (1 - \gamma^*_t) \hat{\mathbf{x}}_{0, e} \big] - \frac{\mathbf{x}_t}{\sigma^2(t)}.
\end{align}
\hfill $\square$

\section{Further Implementation Details}

\textbf{Model Details.}
We leverage the code-base of ADM \cite{2021ADM} to implement the score network $\bm{\mu}_\theta$. The detailed architecture is shown in Tab.~\ref{tab: arch}. The architectures of the Encoder $\mathbf{E}$ and the end-to-end decoder $\mathbf{D}$ are the same with ELIC \cite{2022ELIC}. For the models trained with $\lambda_r \in [0.5, 0.2]$ and $\lambda_r \in [0.1, 0.05, 0.02]$, we load the pre-trained ELIC with $\lambda = 0.004$ and $\lambda = 0.008$ respectively. The sources of the compared methods are given in Tab.~\ref{tab: source}. We thank their owners for their contributions to the community. Following previous works, we pad the images to integral multiple to the patch of $64\times64$ during inference.

\begin{table}[htpb]
    \caption{The sources of the compared methods.}
    \centering
    \begin{tabular}{c|c|c}
    \toprule
         Classification & Method & URL
         \\
         \midrule
         \multirow{1}{*}{MSE-Oriented} & ELIC & \url{https://github.com/VincentChandelier/ELiC-ReImplemetation}
         \\
         \midrule
         \multirow{5}{*}{Perceptual} & \multirow{2}{*}{HiFiC} & \url{https://github.com/Justin-Tan/}
         \\
         & &
         \url{high-fidelity-generative-compression}
         \\
         & CDC & \url{https://github.com/buggyyang/CDC_compression}
         \\
         & \multirow{2}{*}{ILLM} & \url{https://github.com/facebookresearch/}
         \\
         & &
         \url{NeuralCompression/tree/main/projects/illm}
         \\
    \bottomrule
    \end{tabular}
    \label{tab: source}
\end{table}

\textbf{Implementations of other Methods and Metrics.} We implement LPIPS by \url{https://github.com/S-aiueo32/lpips-pytorch/tree/master} and all other metrics by \url{https://github.com/chaofengc/IQA-PyTorch}. When calculating FID, we crop images to $256\times256$ patches. We crop all the images two times from the start position $(0, 0)$ and $(128, 128)$ without overlap. For BPG, we employ BPG v0.9.8 through the following script:

\texttt{\# Encode}

\texttt{bpgenc -o \$binary\_file -q \$qp \$input\_image}

\texttt{\# Decode}

\texttt{bpgdec -o \$recon\_image \$binary\_file}

We leverage quantizer parameters in the set of $[32, 35, 37, 40, 45]$.

\begin{table}[htbp]
    \caption{Detailed architecture of our Model. The model size is the summation of the score network $\bm{\mu}_\theta$, the encoder $\mathbf{E}$, the entropy model and the end-to-end decoder $\mathbf{D}$.}
    \centering
    \begin{tabular}{cc}
        \toprule
         & Entire Model \\
        \midrule 
        Model size & 73.79M
        \\ 
        Channels & 96
        \\
        Depth & 2
        \\
        Channels multiple & 1,1,2,2,3
        \\
        Heads & 4
        \\
        Attention resolution & None
        \\
        BigGAN up/downsample & \checkmark
        \\
        Dropout & 0.0
        \\
        \bottomrule
    \end{tabular}
    \label{tab: arch}
\end{table}

\begin{figure}
    \centering
    \includegraphics[width=\linewidth]{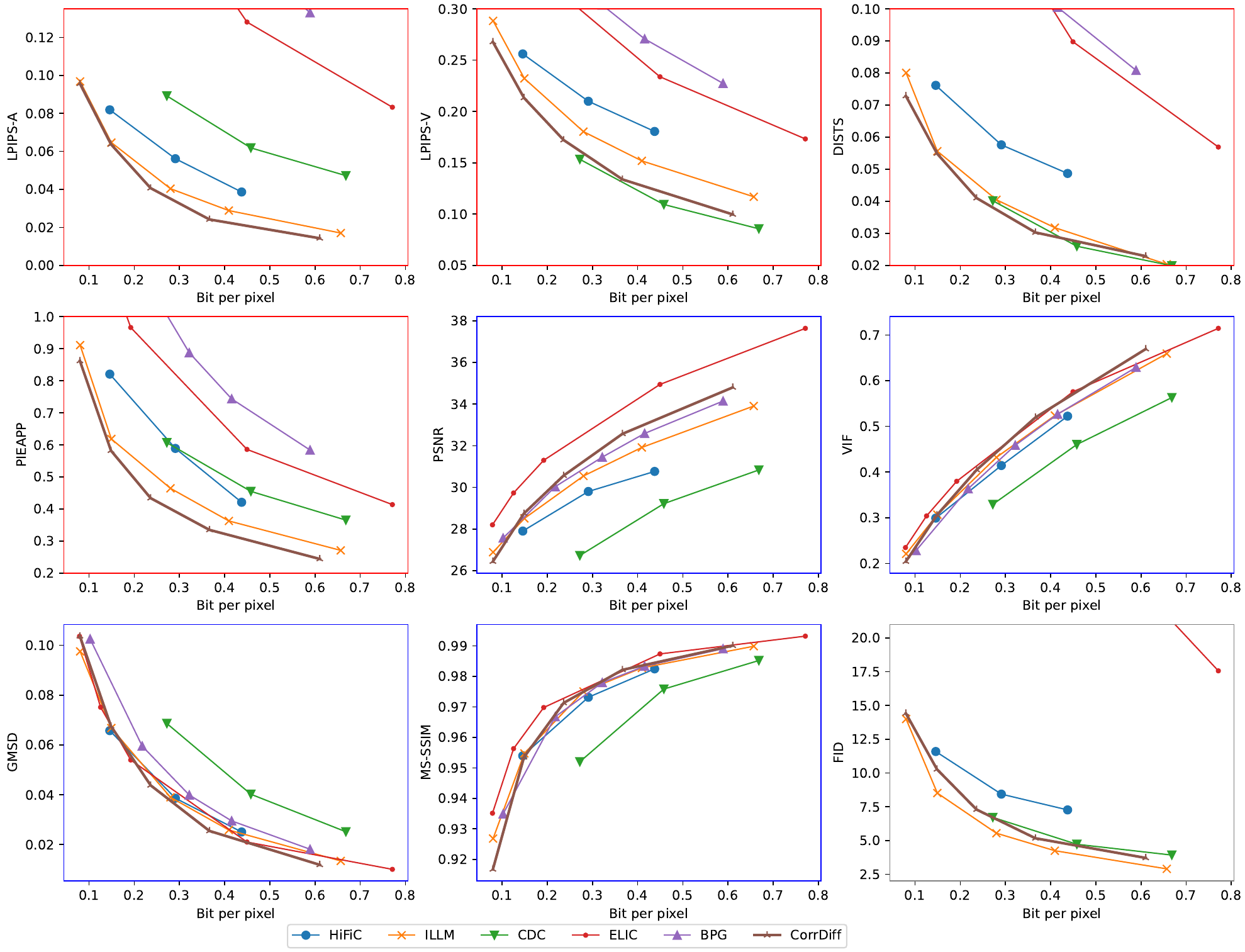}
    \caption{performance of diverse metrics on DIV2K test dataset. \textbf{[Zoom in for best view]}}
    \label{fig: div2k rd}
\end{figure}

\begin{figure}
    \centering
    \includegraphics[width=\linewidth]{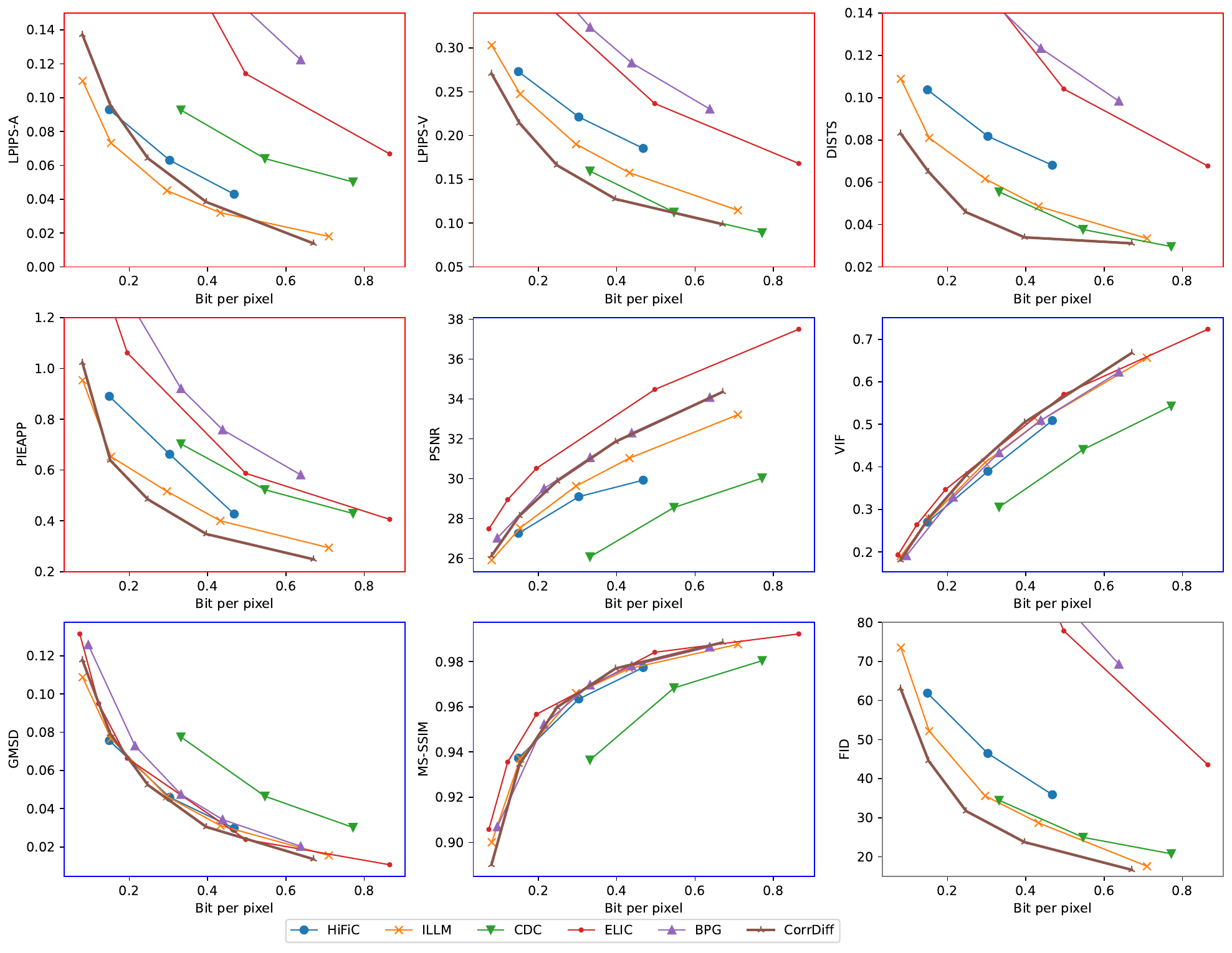}
    \caption{performance of diverse metrics on Kodak dataset. \textbf{[Zoom in for best view]}}
    \label{fig: kodak rd}
\end{figure}

\section{performance on other datasets}
\label{sec: rd on other dataset}

We further show the performance on the dataset of Kodak \cite{1993Kodak} and DIV2K-test \cite{2017DIV2K} in Fig.~\ref{fig: kodak rd} and Fig.~\ref{fig: div2k rd}. The performance on these datasets are similar to the performance on CLIC \textit{professional} which has been shown in the main paper.

\section{Full Versions of Visual Results}
\label{sec: sup visual}

\begin{figure}[b]
    \centering
    \includegraphics[width=0.72\linewidth]{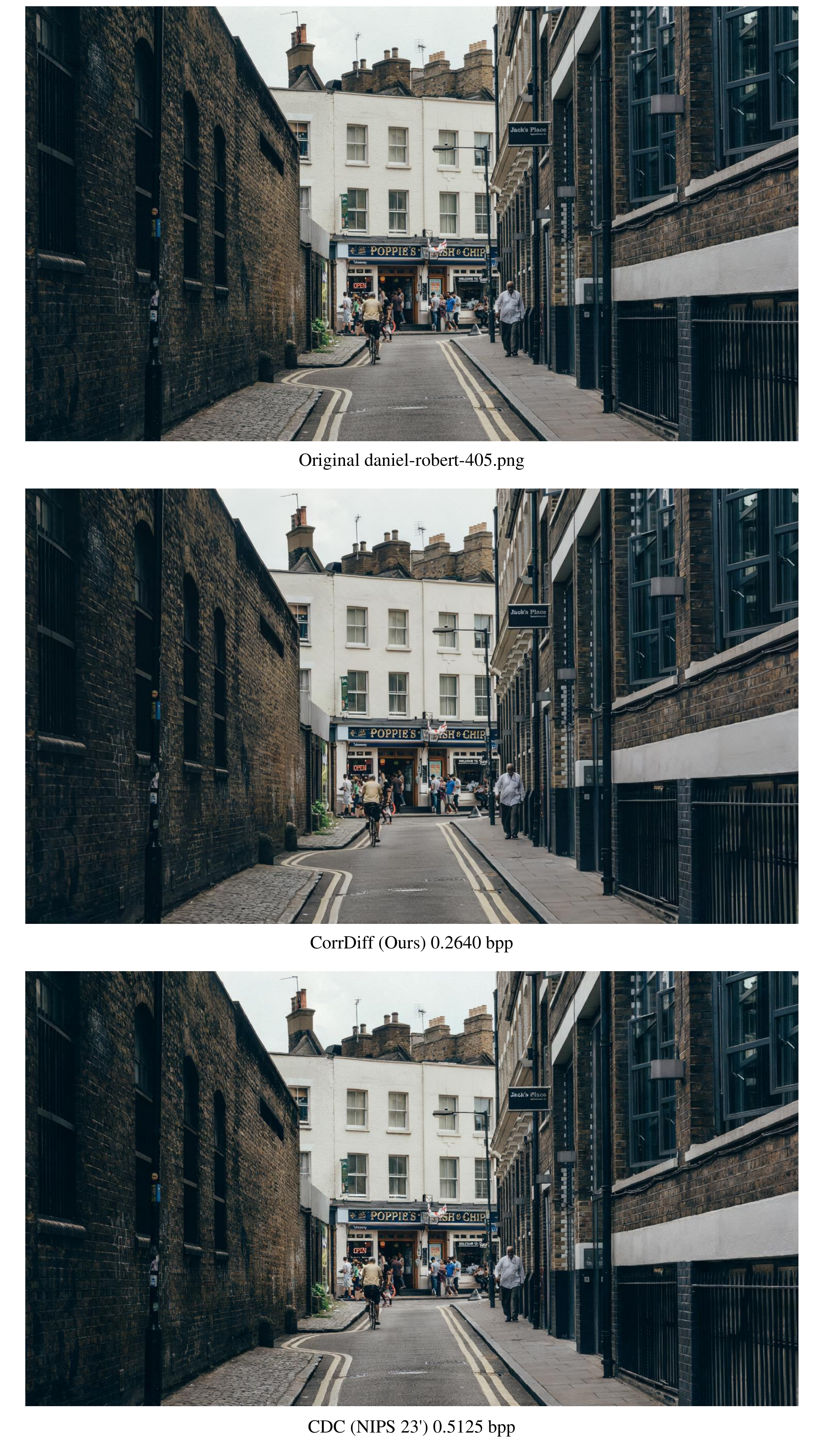}
    \caption{Full version of daniel-robert-405.png from CLIC \textit{professional} dataset. \textbf{[Zoom in for best view]}}
    \label{fig: sup vis 1}
\end{figure}

\begin{figure}
    \centering
    \includegraphics[width=0.72\linewidth]{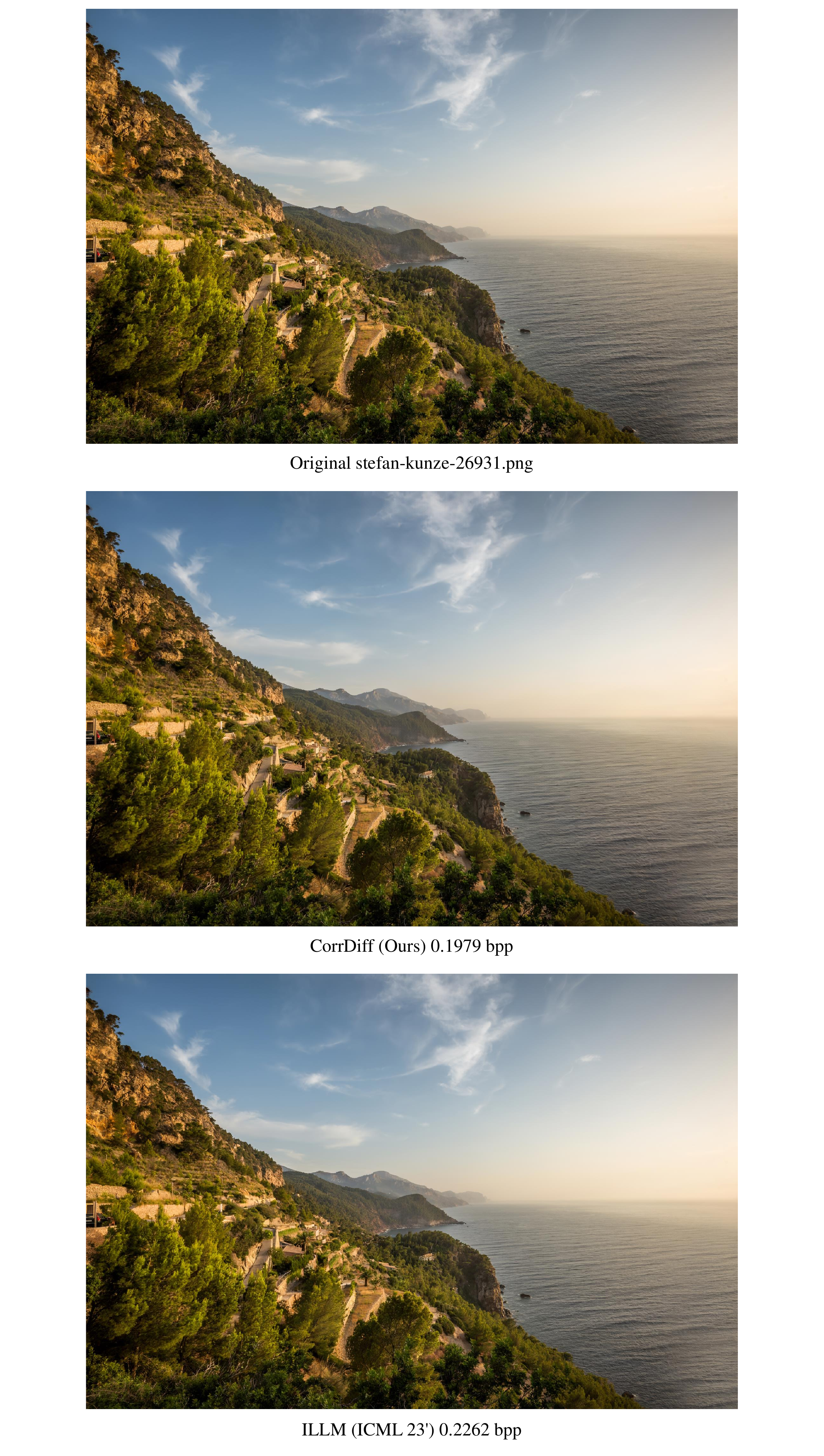}
    \caption{Full version of stefan-kunze-26931.png from CLIC \textit{professional} dataset. \textbf{[Zoom in for best view]}}
    \label{fig: sup vis 2}
\end{figure}

\begin{figure}
    \centering
    \includegraphics[width=0.72\linewidth]{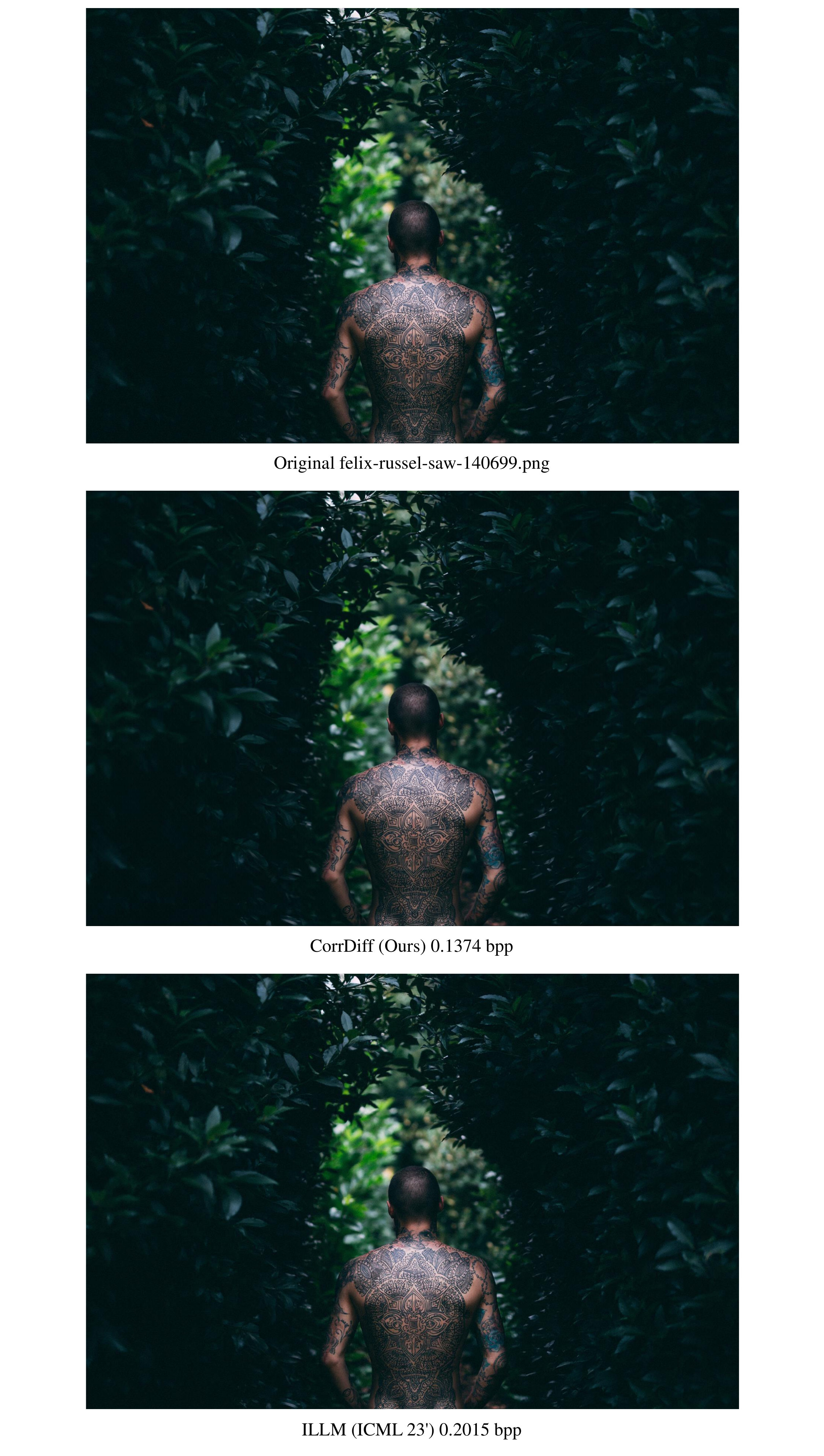}
    \caption{Full version of felix-russel-saw-140699.png from CLIC \textit{professional} dataset. \textbf{[Zoom in for best view]}}
    \label{fig: sup vis 3}
\end{figure}

We show the full versions of the images we have given in the main paper in this section in Fig.~\ref{fig: sup vis 1},~\ref{fig: sup vis 2},~\ref{fig: sup vis 3}.

\end{document}